# Unraveling Intertwined Impacts between Lattice Vacancy and Substrate on Photonic Quasiparticles in Monolayer MoS$_2$

Ning Xu[1†], Daocheng Hong[2†], Xudong Pei[3†], Jian Zhou[1], Fengqiu Wang[1], Peng Wang[3], Yuxi Tian[2]*, Yi Shi[1], Songlin Li[1]*

[1] National Laboratory of Solid-State Microstructures, School of Electronic Science and Engineering, Nanjing University, Nanjing 210023, China
[2] Key Laboratory of Mesoscopic Chemistry of MOE, School of Chemistry and Chemical Engineering, Nanjing University, Nanjing, Jiangsu 210023, China
[3] College of Engineering and Applied Sciences and Jiangsu Key Laboratory of Artificial Functional Materials, Nanjing University, Nanjing 210023, China

† These authors contributed equally to this work.
* Corresponding authors. Emails: sli@nju.edu.cn and tyx@nju.edu.cn.

**ABSTRACT:** Lattice defects and interfacial absorbates represent two extrinsic but ubiquitous factors that exert profound impacts on the luminescent properties of semiconductors. However, their impacts are normally tangled and remain to be separately elucidated. Here, we clarify the individual roles of each factor by tracking the weight evolution of photonic quasiparticles by modulating the densities of sulfur vacancies in monolayer MoS$_2$ via ad hoc defect engineering. In particular, we perform atomistic analyses on the densities of sulfur vacancies by employing atomically resolved scanning transmission electron microscopy to quantitatively characterize the generation rates of sulfur vacancies in MoS$_2$ on two types of substrates with opposite surface hydrophobicity natures. We reveal that the generation rate of sulfur vacancies can even double on the hydrophilic substrates. More importantly, the impact of sulfur vacancies over the weight of trionic emission is closely associated with the substrate hydrophobicity, which is manifested noticeably in the hydrophobic substrates, but insignificant in the hydrophilic ones. The results represent an in-depth understanding on the roles of the extrinsic factors on the luminescent properties in atomically thin semiconductors.

## INTRODUCTION

Two-dimensional (2D) transition-metal dichalcogenides (TMDCs), such as MoS$_2$ and WS$_2$ monolayers, have attracted great interest in advanced electronics, optoelectronics, and photonics due to their unique dimensionality and remarkable luminescent properties.[1–3] In striking contrast to conventional bulk semiconductors, the reduced spatial dimensionality in 2D TMDCs leads to a significant reduction in effective dielectric screening, which gives rise to generally high exciton binding energies on the orders of several hundred meV.[4,5] Such an over-room-temperature energy scale lays the foundation for operating photonic quasiparticles[6,7] in practical applications. However, the high binding energy scale also allows for the generation of multiply bounded quasiparticles, such as charged trions,[8–10] constituting an undesirable factor quenching the luminescent yield, since they are often evolved into the non-radiative recombination pathways. Hence, the physical origins of the charged trions become one of the crucial concerns deserved to be elucidated.

Moreover, due to the full exposure of lattice atoms, the photonic properties of 2D TMDCs are proven exceptionally sensitive to extrinsic factors. Even the background charges arising from the lattice vacancies and substrate charge traps can significantly change the overall luminescent characteristics, including the weights of the bounded quasiparticles, luminescent yields and energy scales.[10–12] In practice, the individual impacts from these two factors are normally tangled and it is difficult to differentiate them. To date, relevant efforts have been rarely reported and, thus, the evolution of quasiparticle weights with the extrinsic factors remains to be well understood.

Here, we employed ad hoc defect engineering and atomically resolved scanning transmission electron microscopy (STEM) to control and characterize the density of atomic vacancies (abbreviated as [V]) in TMDC monolayers. The on-demand modulation of [V] allows us to distinguish the realistic impacts between the lattice vacancies in semiconductors and the charge traps from substrates. Also, the substrate hydrophobicity is identified as an important factor responsible for the charge transfer behavior from underlying substrates. We uncovered the complicated interactions among these factors by tracking the evolution of weights of trions and excitons in the photoluminescent (PL) spectra on substrates with different characteristics in hydrophobicity. In striking contrast to the cases on hydrophobic substrates, such as PDMS and PMMA, where the weight of trion emission roughly increases linearly with [V], the weight variation is almost negligible on the hydrophilic substrates with high charge traps, such as $SiO_2$ and $Al_2O_3$. Finally, the physical origin is briefly discussed.

**RESULTS AND DISCUSSION**

**Ad hoc defect engineering.** The $MoS_2$ monolayers were used as the research platform owing to their direct energy gap and sensitivity of optical response to lattice vacancies. In comparison with traditional bulk materials, the nature of atomic thickness renders them fragile and unendurable for high-energy defect engineering, such as plasma bombardment,[13,14] electron beam,[15,16] chemical etching,[17,18] and heteroatom doping.[19,20] For this reason, we employed a mild and convenient defect engineering strategy by soaking them in hydrogen peroxide $H_2O_2$ to modulate [V] at the surfaces.

To quantify [V] at different stages and accurately analyze the effect of defect engineering, we employed aberration-corrected STEM to examine the lattice conditions and made statistics on [V] over multiple areas and samples.[21–25] At first, the samples were prepared using a wet-transfer method (see Methods for details) and transferred onto STEM grids (Figure 1a). Figure 1b shows the schematic atomic structures for a 2H-$MoS_2$ monolayer viewed from the basal plane and cross-section, where the blue and yellow balls represent the Mo and S atoms, respectively. In the case of exfoliated $MoS_2$ monolayers, sulfur vacancies represent the most common defects because of the lower energy of formation.[26] Figure 1c displays a typical annular dark-field (ADF) STEM image where a sulfur vacancy appears, as indicated by the dotted yellow circle. In such an imaging mode, heavier atoms would exhibit a higher intensity than the lighter ones. As a result, the most and less bright spots in Figure 1c correspond to the Mo and S atoms, respectively. The existence of sulfur vacancies can also be seen in the profile line along the atomic chain, where a lower intensity is observed in the site of vacancy (Figure 1d). In this way, the regions containing vacancy defects can be clearly visualized and counted for statistical purposes. In addition, to minimize the extra bombardment effect during STEM imaging (Figure S2), the exposure time was set at 8 s during imaging acquisition.

Figure 1e depicts the typical atomic images for the $MoS_2$ monolayers defect-engineered for varied times on PDMS surfaces and the trend of vacancy generation. The positions of sulfur vacancies are indicated by yellow circles. For statistical purposes, we collected more than 25 independent regions with an area of $4 \times 4$ nm$^2$ to extract the [V] levels at each defect engineering stage; the corresponding data are given in Figure 1f. In Figure 1g, we further plot [V] versus treatment time ($t$) of defect engineering to estimate the generation rate of [V] with the defect engineering, which reveals a linear relation between them, resulting in a generation rate of $1.5 \times 10^{12}$ min$^{-1}$cm$^{-2}$ and a residual [V] at $2.5 \times 10^{13}$ cm$^{-2}$ in the pristine $MoS_2$. The relatively low generation rate indicates that oxidant soaking is a convenient and controllable way for ad hoc defect engineering.

**Evolution of trionic emission weight with [V].** Figure 2a shows the typical optical image for an as-exfoliated $MoS_2$ sample with a local monolayer (1L) area on a PDMS substrate. We tracked the evolution of PL spectra with increasing [V] from 2.5 to 7.1×10$^{13}$ cm$^{-2}$ (Figure 2b) and focused on the regime of the low-energy exciton around 1.7-2.1 eV, where the most characteristic PL emissions from the charged trions (T) and neutral excitons (X) are located. The features including intensities and positions of the excitonic and trionic emissions were analyzed and extracted by Lorentzian fittings. It is found that the weight of trionic emission increases with [V] and a roughly linear dependence can be seen in the plot of intensity ($I$) ratio $I(T)/[I(X)+I(T)]$ (Figure 2c). Accordingly, the weight of excitonic emission is reduced at high [V] levels (inset of Figure 2c), due to the additional doping effect from vacancies.

According to the mass action model,[27] the ratio of $I(T)/[I(X)+I(T)]$ is ruled by

$$\frac{I(T)}{I(X)+I(T)} = \frac{\Gamma_T N_T}{\Gamma_X N_X + \Gamma_T N_T} \quad (1)$$

where $\Gamma_T$ and $\Gamma_X$ are the radiative recombination rates of trions and excitons, and $N_T$ and $N_X$ are the densities of trions and excitons. The $\Gamma_T$ and $\Gamma_X$ are reported to be influenced by the environmental dielectrics, that is $\Gamma_T/\Gamma_X \propto (\kappa_{eff})^\delta$, where $\kappa_{eff}$ is the effective relative dielectric constant and $\delta$ is a material relevant coefficient. By assuming fixed values of $\kappa_{eff}$ and $\delta$ within the highest experimental [V] level (1.1×10$^{14}$ cm$^{-2}$), the ratio of $\Gamma_T/\Gamma_X$ is then fixed and thus, the change of the ratio of $I(T)/[I(X)+I(T)]$ is only associated with the variations of $N_T$ and $N_X$. Furthermore, the ratio of $I(X)/I(T)$ can be expressed as[28]

$$\frac{I(X)}{I(T)} \propto \frac{1}{n_e} e^{\frac{\varepsilon_T}{k_B T}} \quad (2)$$

where $n_e$ is density of background charges, $\varepsilon_T$ is the binding energy of trions (Figure S3), and $k_B T$ is the thermal energy at room temperature. Hence, it is inferred that the enhanced ratio of $I(T)/[I(X)+I(T)]$ (increased $N_T$ and decreased $N_X$) with increasing [V] is a direct consequence of enhancement of background charges in the MoS$_2$ monolayers. As [V] increases, redundant electrons are produced due to the presence of dangling bonds of molybdenum atoms that were originally bonded with sulfur atoms (Figure 2d). Thus, a certain number of them recombine with the neutral excitons and form charged trions through complicated Coulomb interactions.

In previous studies, it has been reported that the formation of a small number of sulfur vacancies is favorable for suppression the density of charged trions and results in a dominant excitonic emission in the PL spectra,[14,20] through various doping compensation mechanisms, including chemical bonding or physical absorption of O$_2$ and H$_2$O molecules.[29] This phenomenon is absent in our experiment, likely because of the higher [V] levels achieved with defect engineering, which results in a higher proportion of unfilled vacancies or weak physical adsorption of ambient oxygen, where the charge transfer from oxygen to MoS$_2$ is as low as 0.02e per atom.[14] Overall, the doping compensation effect was not clearly observed in our experiment.

Time-resolved PL lifetime measurements for samples at different [V] levels were also carried out to identify the effect of [V] on the quasiparticle lifetimes (Figure 2e). An overall decaying trend is seen for the lifetime as [V] increases (Figure 2f), which can be attributed to the reduction in the population of radiative excitons due to the increase of defect-mediated non-radiative recombination probabilities. As previously reported by Lien *et al.*,[3,30] the excitonic recombination is totally radiative and the trionic emission is mainly non-radiative at low generation rates. Hence, our observation of the defect-assisted decay in lifetime can be understood as the result of depression of excitonic emission and the increase of trionic emission due to the increase in the proportion of background charges produced by the vacancies.

**Effect of substrate hydrophobicity.** To further shed light on the effect of charge traps in underlying substrates, we also extended the experiment to versatile substrates. Since most charge traps are closely associated with absorbed humidity, we adopted other three substrates with different hydrophobicity natures, in addition to the hydrophobic PDMS substrates. Among the three, a hydrophobic substrate PMMA (analogous to PDMS) and two hydrophilic substrates (i.e., Al$_2$O$_3$ and SiO$_2$) were selected.[31–34]

Before experiment, the generation rates of [V] ($r_{[V]}$) of monolayer MoS$_2$ supported on these substrates were carefully checked to verify the possible variation in reactive activity.[35] Figure 3a shows the statistical [V] data for $MoS_2$ samples with different values of defect engineering time ($t$) on the hydrophilic $SiO_2$ substrates. The overall trend is plotted in Figure 3b, with extracting a higher $r_{[V]}$ value of ~$2.8\times10^{12}$ cm$^{-2}$ min$^{-1}$ than that on the hydrophobic PDMS substrate. Besides $SiO_2$ and PDMS, the values of $r_{[V]}$ on the PMMA and $Al_2O_3$ substrates were also explored, as shown in Figure 3c (see Figure S4 and S5 for details). Likewise, $r_{[V]}$ is higher on the hydrophilic $Al_2O_3$ substrate (~$2.8\times10^{12}$ cm$^{-2}$ min$^{-1}$) than the hydrophobic PMMA (~$1.5\times10^{12}$ cm$^{-2}$ min$^{-1}$). In general, we find the values of $r_{[V]}$ can be categorized into two broad types: a higher $r_{[V]}$ around $2.8\times10^{12}$ cm$^{-2}$ min$^{-1}$ for the hydrophilic substrates and a lower $r_{[V]}$ around $1.5\times10^{12}$ cm$^{-2}$ min$^{-1}$ for the hydrophobic substrates. Thus, the vacancy generation rate roughly doubles for $MoS_2$ on the moisture surroundings, as compared with the dry condition.

The accurate estimation of $r_{[V]}$ allows us to correctly analyze the evolution of quasiparticle weights under different conditions. We also transferred $MoS_2$ monolayers onto the three substrates (Figure 4a-c) to further collect their PL spectra. Figure 4d-f illustrates the dynamic evolution of PL spectra with [V]. As performed above, the weights of trionic and excitonic emissions are analyzed with Gaussian fittings. Figure 4g-i summarizes the corresponding dynamic evolution of trionic emission weight (i.e., $I(T)/[I(X)+I(T)]$) with [V] for samples on PMMA, $Al_2O_3$, $SiO_2$, respectively, with the insets showing the complementary weights of excitonic emission. As can be seen, the trend of trionic weight on PMMA is quite similar to that on PDMS, in which the ratio of trionic emission increases (from ~0.2 to ~0.4) linearly with [V] within the experimental uncertainty. The linear fit results in an intercept of 0.1 at zero [V], implying a small weight for trionic emission in the perfect $MoS_2$ on PMMA. We deduce that such a residual trionic weight arises likely from the ambient absorbates and the weak charge transfer from the underlying PMMA. The overall behavior is basically same to the case of PDMS.

By contrast, in case of the hydrophilic substrates, the trionic weights exhibit weak dependence on [V]. As [V] increases, the ratios of $I(T)/[I(X)+I(T)]$ are nearly fixed at 0.4 and 0.2 for $Al_2O_3$ and $SiO_2$, respectively, suggesting the existence of strong interaction between the charge traps and $MoS_2$. It is highly likely that the introduced atomic vacancies can be quickly filled with the interfacial absorbates and the charge components in $MoS_2$ are nearly unchanged. This observation highlights the importance of the substrate absorbates and traps on the luminescent properties of the supported semiconductors.

**Untangling intertwined impacts between substrates and vacancies.** To further look insight into the individual impacts between substrates and vacancies, we then analyze the dependence of trionic weight and PL intensity on [V]. To elucidate the effect of substrate moisture, we summarize in Figure 5a the dependence of trionic weight on [V], i.e., the slope of $I(T)/[I(X)+I(T)]$ versus [V], for all four substrates involved. In general, the slopes for the hydrophilic substrates are one order higher in magnitude than the hydrophobic counterparts. Figure 5b schematically depicts the origins of photonic quasiparticles and relevant recombination pathways, where most photogenerated electron-hole pairs are bound by Coulomb interaction to form excitons.

Since the formation of charged trions is normally attributed to the unbalanced electric neutrality from the background charges in semiconductors. We then discuss the distinct dependence of trionic weights mainly based on the relations with the background charges. Evidently, in our experiment there are two main sources for trionic generation, that is, from the presence of vacancies or substrate interactions. These origins are also consistent with the reports on the mechanisms of the background charges.[10] For simplicity, we then distinguish the origins of trions by abbreviating the substrate and vacancy relevant trions as S- and V-trions, respectively.

On hydrophobic substrates (i.e., PMMA and PDMS), the density of S-trions is initially low at low [V]. However, the V-trions grow quickly with increasing [V], accompanied with the fast decay in PL intensity (Figure 5c). This observation indicates that, besides generation of V-trions, the lattice vacancies also serve as dominant non-radiative recombination centers. In striking contrast, even at low [V], it is likely that the substrate moisture on the hydrophilic substrates initially serves as the non-radiative recombination centers and causes a large number of S-trions due to charge transfer. Thus, the additional vacancies have little influence on the two parameters, resulting in insignificant changes in both the trionic weight and PL intensity. Figure 5d shows schematic diagrams for the trionic components on the two types of substrates. In a word, the absorbed interfacial moisture and relevant species interact more actively with the semiconductors, which highlights the crucial role of interface cleanliness on the overall luminescent properties.

**CONCLUSIONS**

We performed combined atomistic and spectral analyses on the monolayer $MoS_2$ supported by different substrates with opposite hydrophobicity natures to investigate the intertwined effects of lattice vacancy and interfacial absorbates on the photonic quasiparticles in semiconductors. By utilizing the density of lattice vacancies and substrate hydrophobicity as two variant parameters, we unraveled the individual impacts between the two factors. In specific, the charge traps from substrates are mainly responsible for the trionic formation in less defective $MoS_2$ on hydrophilic substrates and the contribution from the lattice vacancies is insignificant even at high [V] levels, while the lattice vacancies become the main origin for the trionic formation on hydrophobic substrates. These results help to detangle the intertwined effects of substrates and lattice vacancies and to disclose the physical origins of the various photonic quasiparticles in semiconductors.

**EXPERIMENTAL METHODS**

**Mechanical exfoliation.** All monolayer $WS_2$ flakes were mechanically exfoliated from synthesized crystals with Scotch tapes. The thickness of monolayer $MoS_2$ was confirmed by PL characteristics, which arises from the unique indirect-to-direct bandgap transition for the TMDC monolayers. The monolayer $MoS_2$ is transferred onto other substrates using PDMS as transfer media.

**Defect engineering.** 30% hydrogen peroxide $H_2O_2$ solutions were employed as oxidant for monolayer $MoS_2$ for creating the sulfur vacancies during defect engineering. The exfoliated monolayer samples were soaked in the $H_2O_2$ solution at room temperature. The different soaking time was used as a parameter to control the density of sulfur vacancies.

**Preparation of STEM specimen.** To investigate the vacancy generation rates of $MoS_2$ monolayers supported on the $SiO_2$, $Al_2O_3$, PDMS and PMMA surfaces, the samples defect-engineered for various times should be wet-transferred from the different surfaces to copper STEM grids before STEM imaging. In case of the soluble and soft PMMA surfaces, the $MoS_2$ samples can be directly detached, with contacted PMMA as the wet-transfer support, from the underlying $SiO_2$/Si substrates after soaking in saturated NaOH solution for 5 min. After proper rinse in deionized water, the $MoS_2$/PMMA bilayers were then transferred to the STEM grids by gently pressing and post-baking at 140 °C for 10 min to remove the water. Finally, the wet-transfer support PMMA was dissolved with acetone and the STEM grid supported $MoS_2$ samples were achieved.

In case of the hard $SiO_2$ (or $Al_2O_3$) surfaces, the $MoS_2$ samples should be first detached from the hard surfaces and transferred onto soluble and soft wet-transfer media (e.g. PMMA polymers). Thus, we first prepared PMMA polymers directly on the $MoS_2$-contained hard surfaces. A slow rotation rate of 2000 rpm and twice spin-coating were employed to increase the thickness and mechanical strength of PMMA. After spin-coating, the entire stacks were then placed at room temperature for 5 hours to cure PMMA to achieve the PMMA/$MoS_2$/$SiO_2$ (or $Al_2O_3$/$SiO_2$)/Si structures. The following transfer procedures were analogous to the case of PMMA mentioned above. In case of the hydrophobic and insoluble PDMS surfaces, the engineered $MoS_2$ monolayers were directly transferred onto the $SiO_2$/Si substrates by soft mechanical pressing and the sequent procedures were similar to the case of $SiO_2$ surface mentioned above.

**HR-STEM imaging.** High-resolution ADF-STEM images were acquired on a double aberration-corrected FEI Titan Cubed G2 60-300 S/TEM at 60 kV. To enhance the contrast of the sulfur sublattices, a medium-range ADF mode was selected by adjusting the camera length properly. The probe current is set at 56 pA and the integration time is 10 s for collecting an image.

**PL and quasiparticle lifetime characterization.** The PL spectra were characterized on a home-made facility using a solid laser with a wavelength of 488 nm operated at 630 W/cm$^2$ as the excitation source. The PL signals were collected by a dry objective lens (Olympus LUCplanFI 40×, NA=0.6) and detected by an sCMOS camera (Andor Sona) after passing through a 473 nm long-pass filter (BLPO1-473-R-25, Semerock).

The quasiparticle lifetime measurements were performed on another home-built system based on Olympus IX73 wide-field microscope and a super-continuous laser (Fianium SC-400). The excitation wavelength and repetition rate employed were respectively 450 nm and 40 MHz. The lifetime spectra were recorded by a time-correlated single-photon counting system (TCSPS, Picoharp 300).

**Supporting Information**
STEM image filtering, effect of electron bombardment during STEM imaging, binding energy of trions; and statistics of [V] in defect engineered $MoS_2$ on PMMA and $Al_2O_3$ surfaces.

**Funding Sources**
This work was supported by the National Natural Science Foundation of China (61974060 and 61674080), the National Key R&D Program of China (2017YFA0206304).

**Competing interests**
The authors declare no competing financial interest.

**Figure captions**

**Figure 1: Quantifying [V] in defect engineered MoS$_2$ monolayers with atomically resolved STEM. a**, A typical image for a MoS$_2$ monolayer on an STEM grid. Scale bar: 2 μm. **b**, Atomic structure for 2H phase MoS$_2$ viewed from the basal and cross-sectional angles. Blue: Mo atoms; golden: S atoms. **c**, A atomically resolved local area. Dashed circle: a sulfur vacancy. Scale bar: 0.5 nm. **d**, Intensity profile along an atomic chain in **c** (blue rectangle). The relatively weak intensity indicates the presence of a sulfur vacancy. **e**, Evolution of lattice vacancies (marked by golden circles) on PDMS surface with various H$_2$O$_2$ treatment times ($t$s). Scale bar: 2 nm. **f**, Histograms of [V] at various $t$s. **g**, [V] as a function of $t$. A linear fit to the [V]-$t$ relation reveals a $r_{[V]}$ of $1.45\times10^{12}$ min$^{-1}$cm$^{-2}$ and an intercept of $2.5\times10^{13}$ cm$^{-2}$.

**Figure 2: Evolution of PL spectra with increasing [V]. a,** Optical image for an as-exfoliated MoS$_2$ sample on PDMS surface. Scale bar: 20 μm. **b**, Normalized PL spectra for monolayer MoS$_2$ treated by H$_2$O$_2$ solution for different $t$s at 0, 10, 20, and 30 min, respectively. The excitonic (X) and trionic (T) emissions are fitted with Lorentzian functions. **c**, The extracted trionic emission weight, $I(T)/[I(X)+I(T)]$, versus [V]. Inset: excitonic emission weight. **d**, Schematic diagrams for trion formation due to the introduction of sulfur vacancies and background charges. **e**, Time-resolved PL lifetime spectra at different [V] levels. **f**, Extracted PL lifetime versus [V].

**Figure 3: Generation rates of [V] ($r_{[V]}$) on surfaces other than PDMS**. **a**, Histograms of [V] on SiO$_2$ surfaces at various $t$s. **b**, [V] versus $t$ where the linear fit reveals a $r_{[V]}$ of $2.8\times10^{12}$ min$^{-1}$cm$^{-2}$ and a residual [V] of $2.2\times10^{13}$ cm$^{-2}$ for pristine MoS$_2$. **c**, Summary of $r_{[V]}$ on the four types of employed surfaces, where the hydrophobic surfaces generally exhibit a low $r_{[V]}$ and the hydrophilic surfaces show a high $r_{[V]}$.

**Figure 4: Evolution of quasiparticle emission weights on surfaces other than PDMS.** Optical images of MoS$_2$ samples transferred on (**a**) PMMA, (**b**) Al$_2$O$_3$ (**c**) SiO$_2$ surfaces. Scale bar: 20 μm. (**d-f**) Corresponding normalized PL spectra and (**g-i**) Emission weights from trions and excitons at different [V] levels.

**Figure 5: Unraveling intertwined effects of substrate and lattice vacancy on the photonic quasiparticles in MoS$_2$ monolayers. a**, Dependence of trionic weight on [V], i.e., the slope of $I(T)/[I(X)+I(T)]$ versus [V], for all four surfaces involved. **b**, Recombination pathways for excitons and trions in MoS$_2$ monolayers. **c**, PL intensity versus [V] for MoS$_2$ monolayers on the four surfaces. **d**, Schematic diagrams for the substrate- and vacancy-relevant trionic components formed on the two types of surfaces.

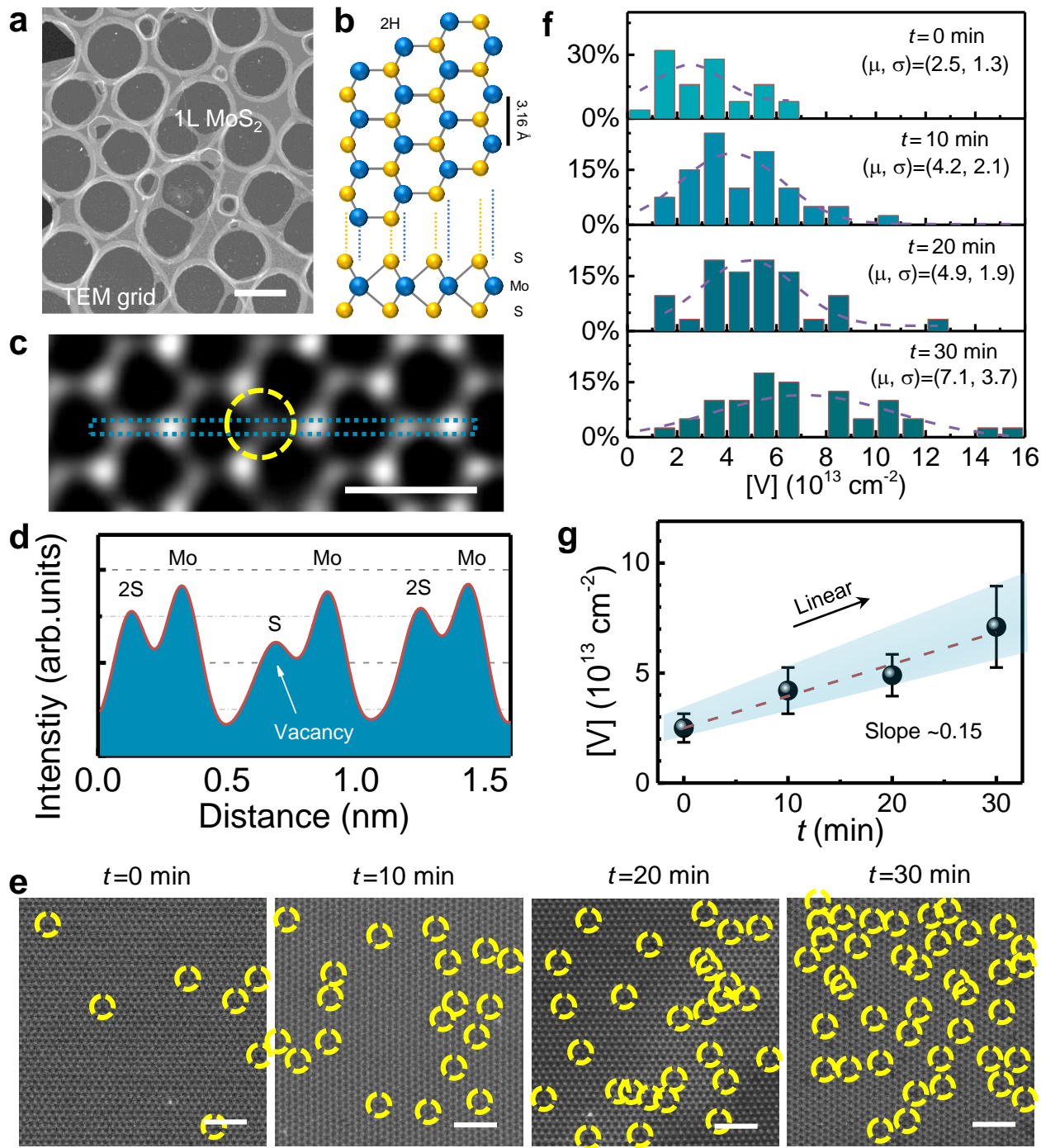

Figure 1

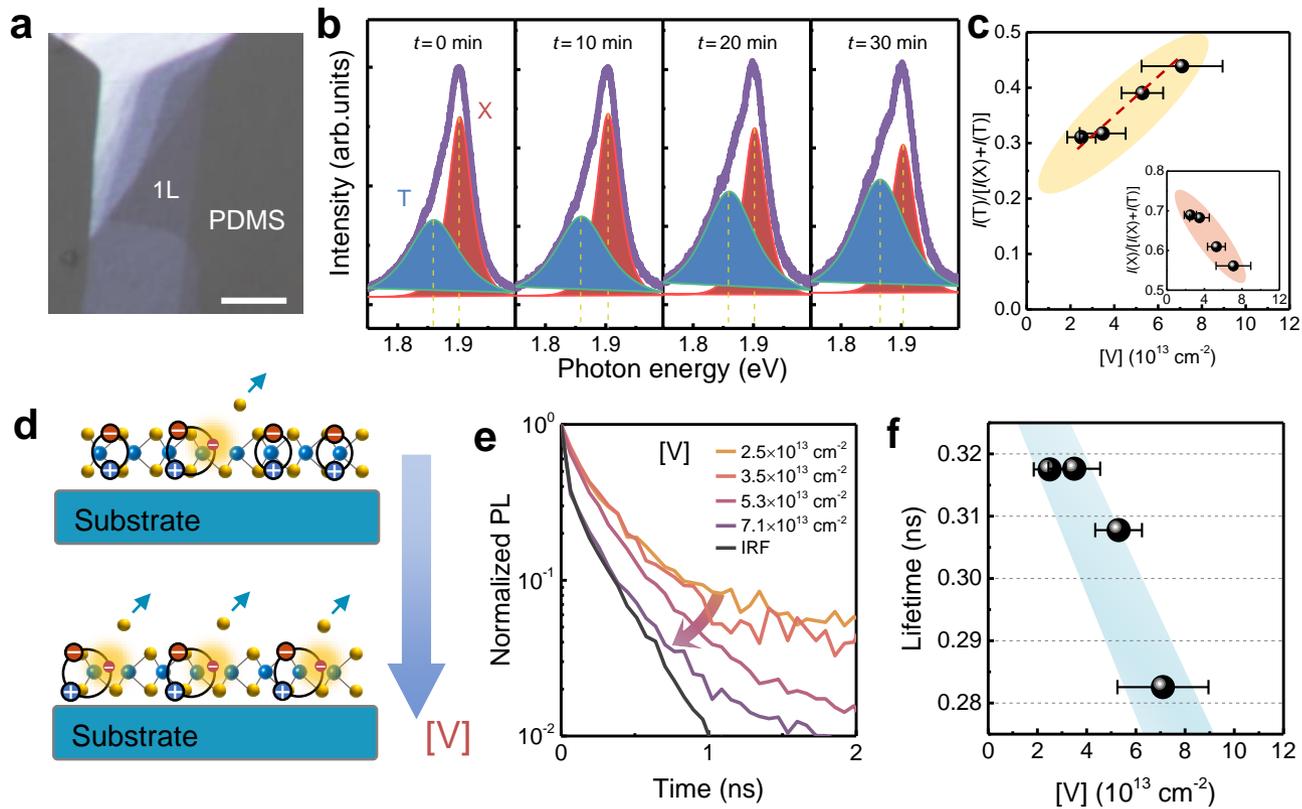

Figure 2

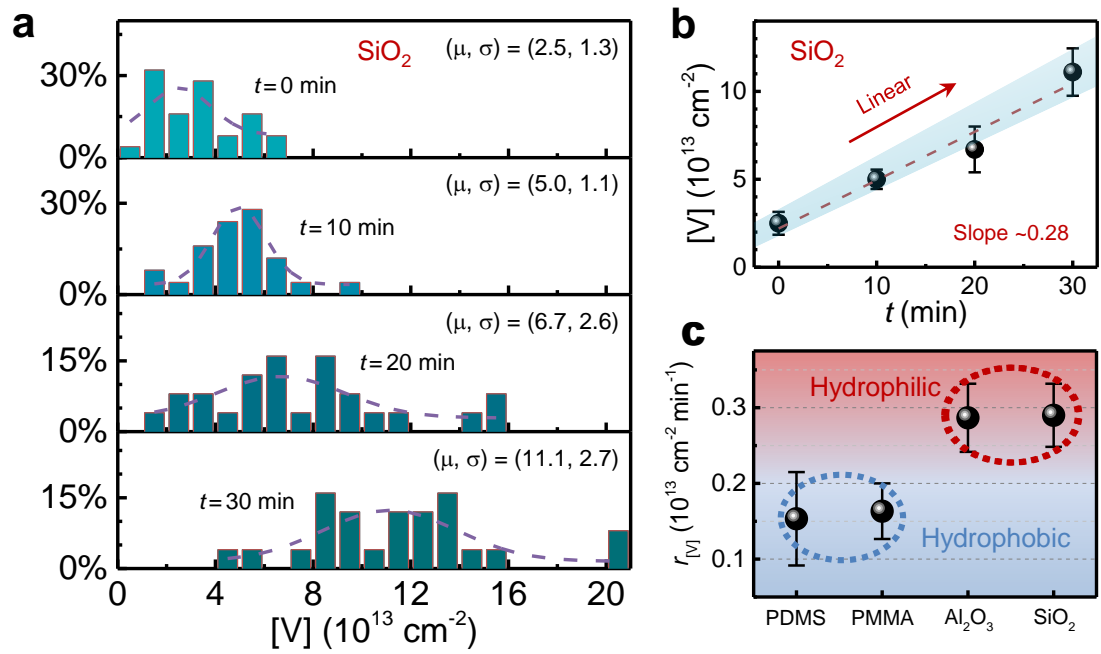

Figure 3

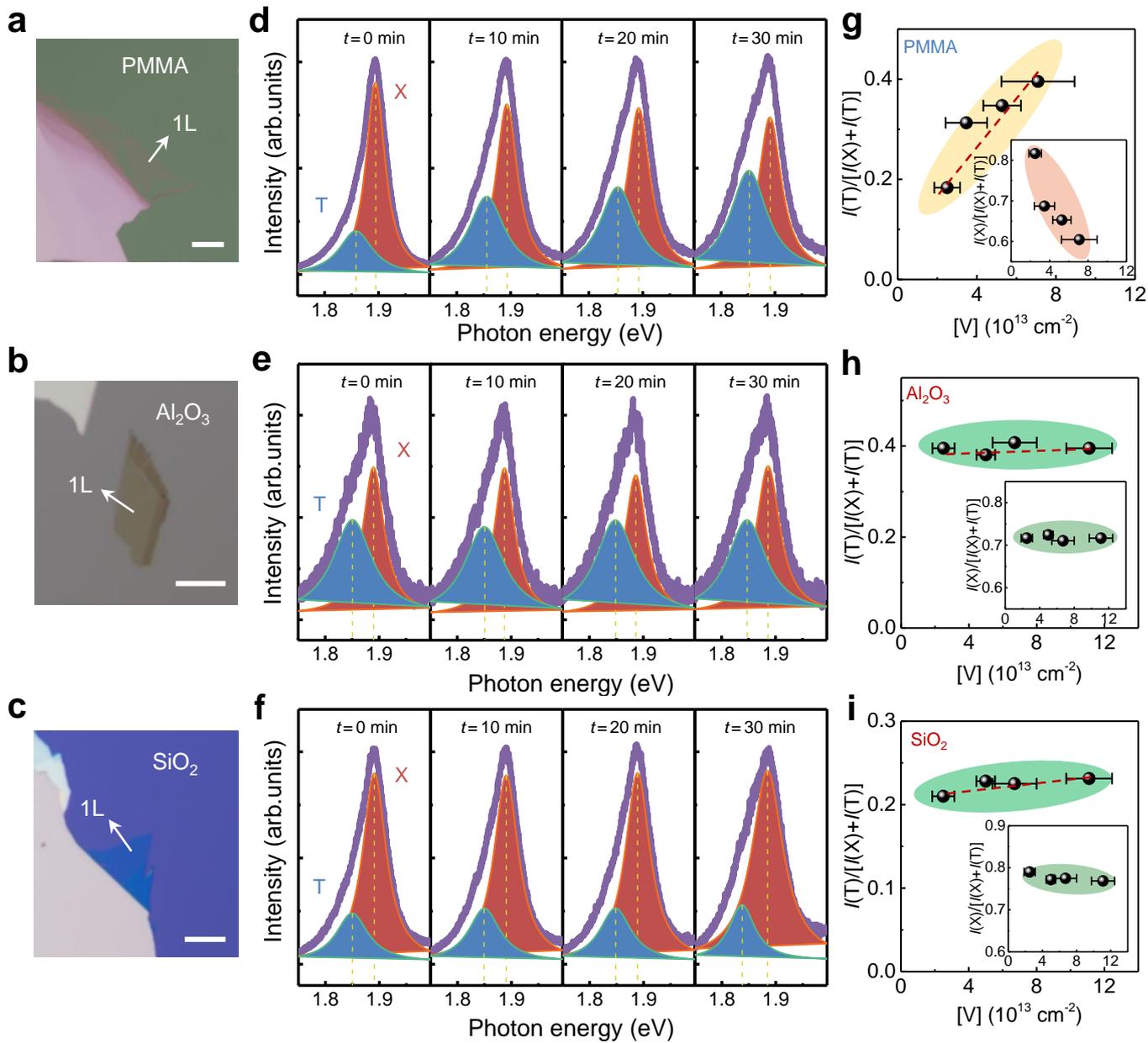

Figure 4

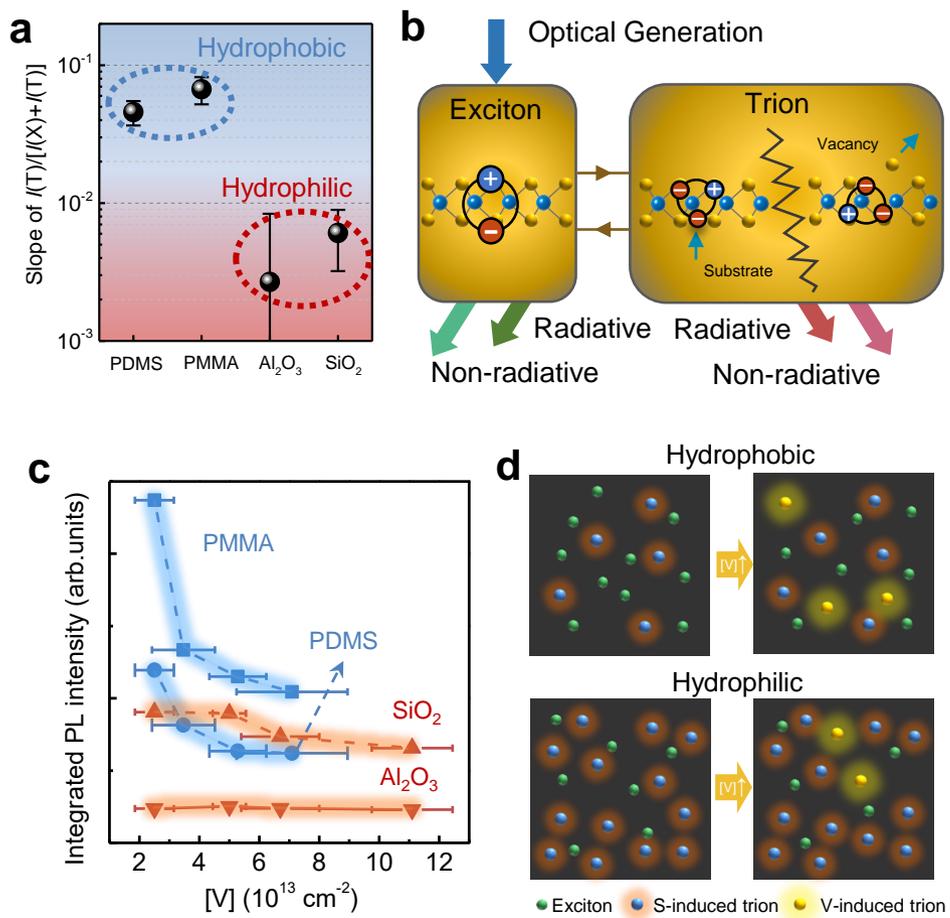

Figure 5

*Supporting Information*

# Unraveling Intertwined Impacts between the Lattice Vacancy and Substrate on Photonic Quasiparticles in Monolayer MoS$_2$

Ning Xu[1†], Daocheng Hong[2†], Xudong Pei[3†], Jian Zhou[1], Fengqiu Wang[1], Peng Wang[3], Yuxi Tian[2]*, Yi Shi[1], Songlin Li[1]*

[1] National Laboratory of Solid-State Microstructures, School of Electronic Science and Engineering, Nanjing University, Nanjing 210023, China
[2] Key Laboratory of Mesoscopic Chemistry of MOE, School of Chemistry and Chemical Engineering, Nanjing University, Nanjing, Jiangsu 210023, China
[3] College of Engineering and Applied Sciences and Jiangsu Key Laboratory of Artificial Functional Materials, Nanjing University, Nanjing 210023, China

† These authors contributed equally to this work.
* Corresponding authors. Emails: sli@nju.edu.cn and tyx@nju.edu.cn.

**Contents**

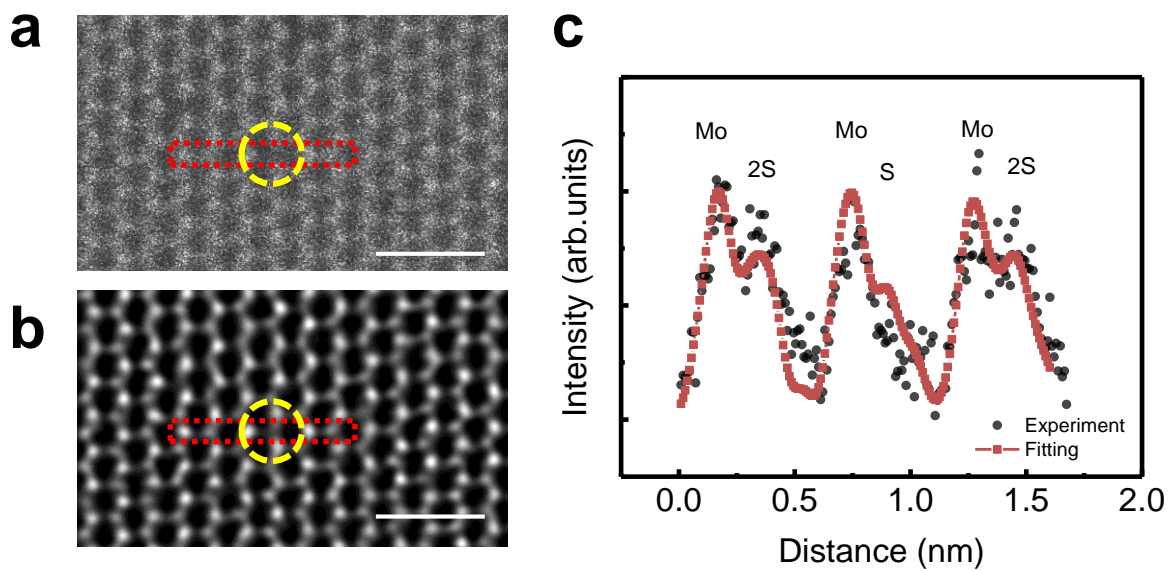

**Figure S1. STEM-image processing by the Wiener and average background subtracted filtering**. **a**, A typical pristine ADF-STEM image. **b**, Corresponding filtered image. Scale bar: 1 nm. **c**, Intensity profiles along the atomic chains indicated in the pristine and filtered images. The profile line becomes smoother after noise filtering. Hence, by examining the atomic sites with reduced intensity, the sulfur vacancies can be promptly identified.

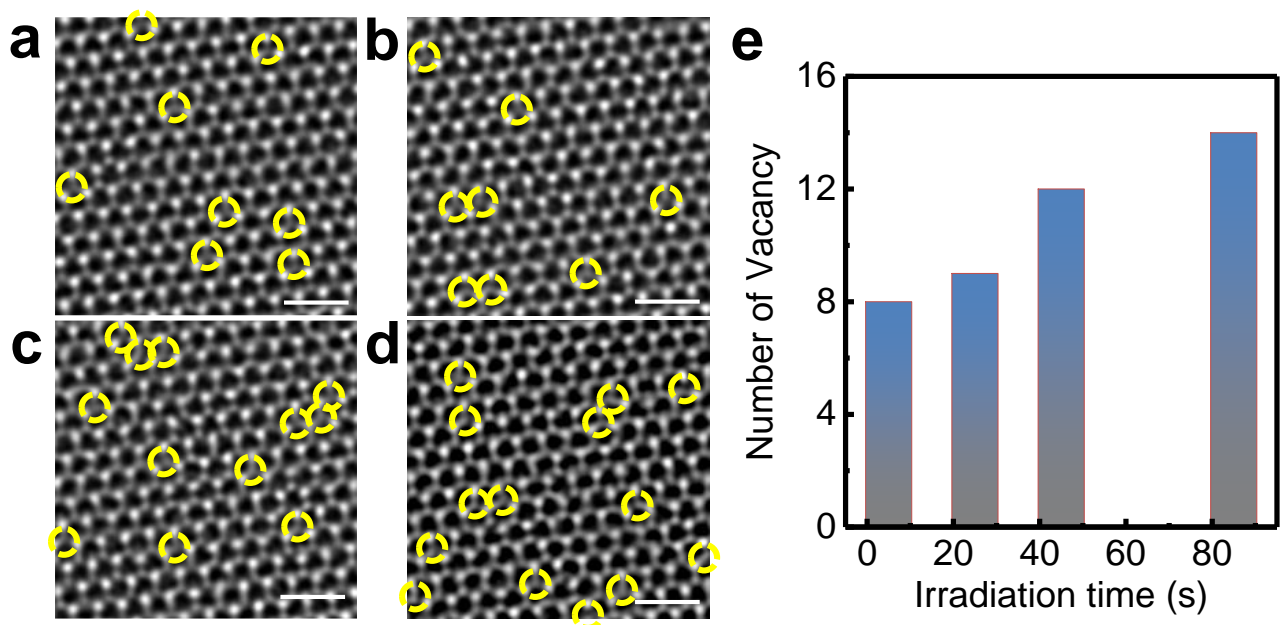

**Figure S2. Effect of electron bombardment on the number of sulfur vacancies in lattices. a**–**d**, ADF-STEM images taken at different irradiation times. The probe current is set at 56 pA and the integration time is 10 s for collecting an image. The imaging area is 4×4 nm$^2$ for all images, where individual sulfur vacancies are denoted by dashed yellow circles. **e**, Evolution of the number of lattice vacancies with increasing beam irradiation time, where a roughly linear correlation is observed.

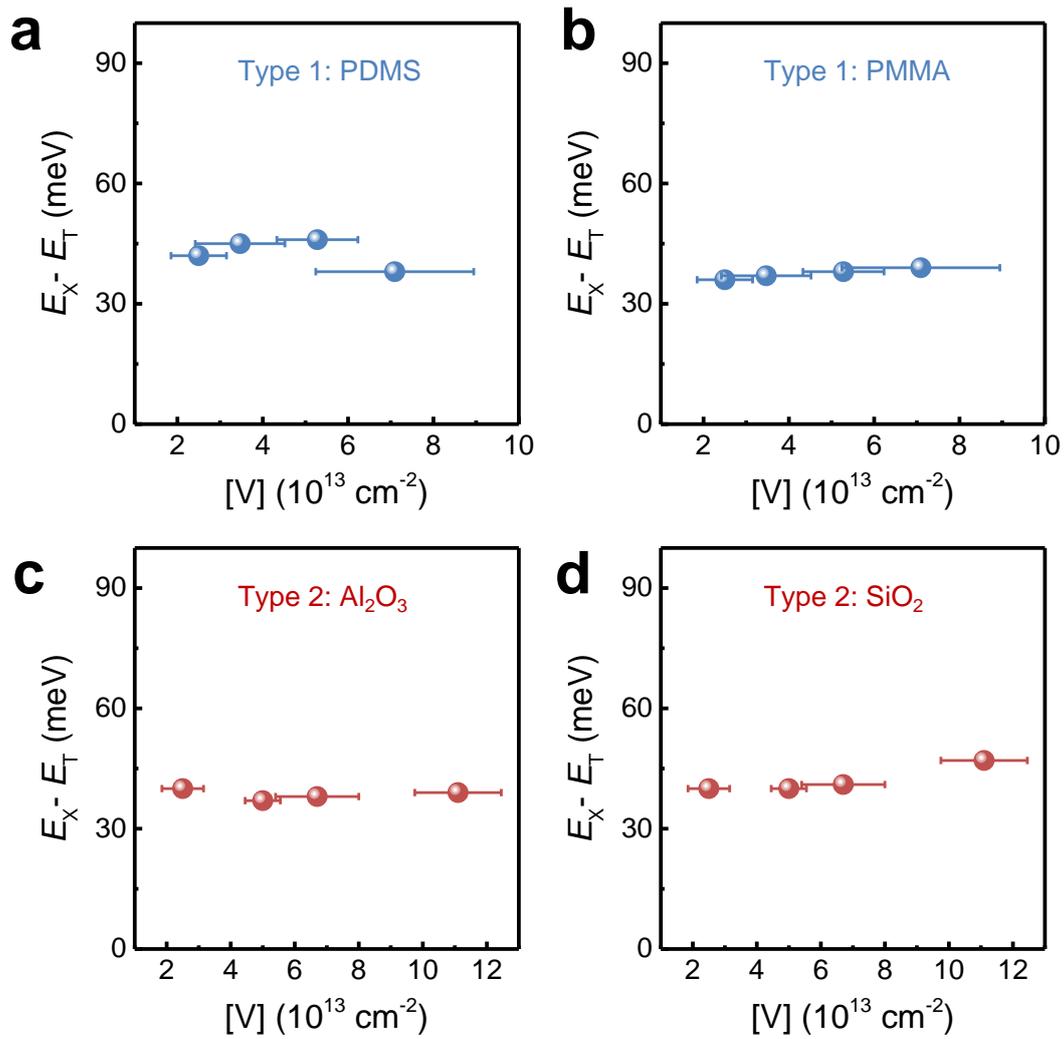

**Figure S3. Trion binding energy ($E_T^b$) versus [V] on different substrates.** (a) PDMS, (b) PMMA, (c) $Al_2O_3$ and (d) $SiO_2$. $E_T^b$ is defined as the difference between the positions of the exciton and trion peaks, that is $E_X - E_T$, which indicates the minimum energy required to remove one electron from a trion. $E_T^b$ shows negligible dependence on [V] up to $1.2 \times 10^{14}$ cm$^{-2}$.

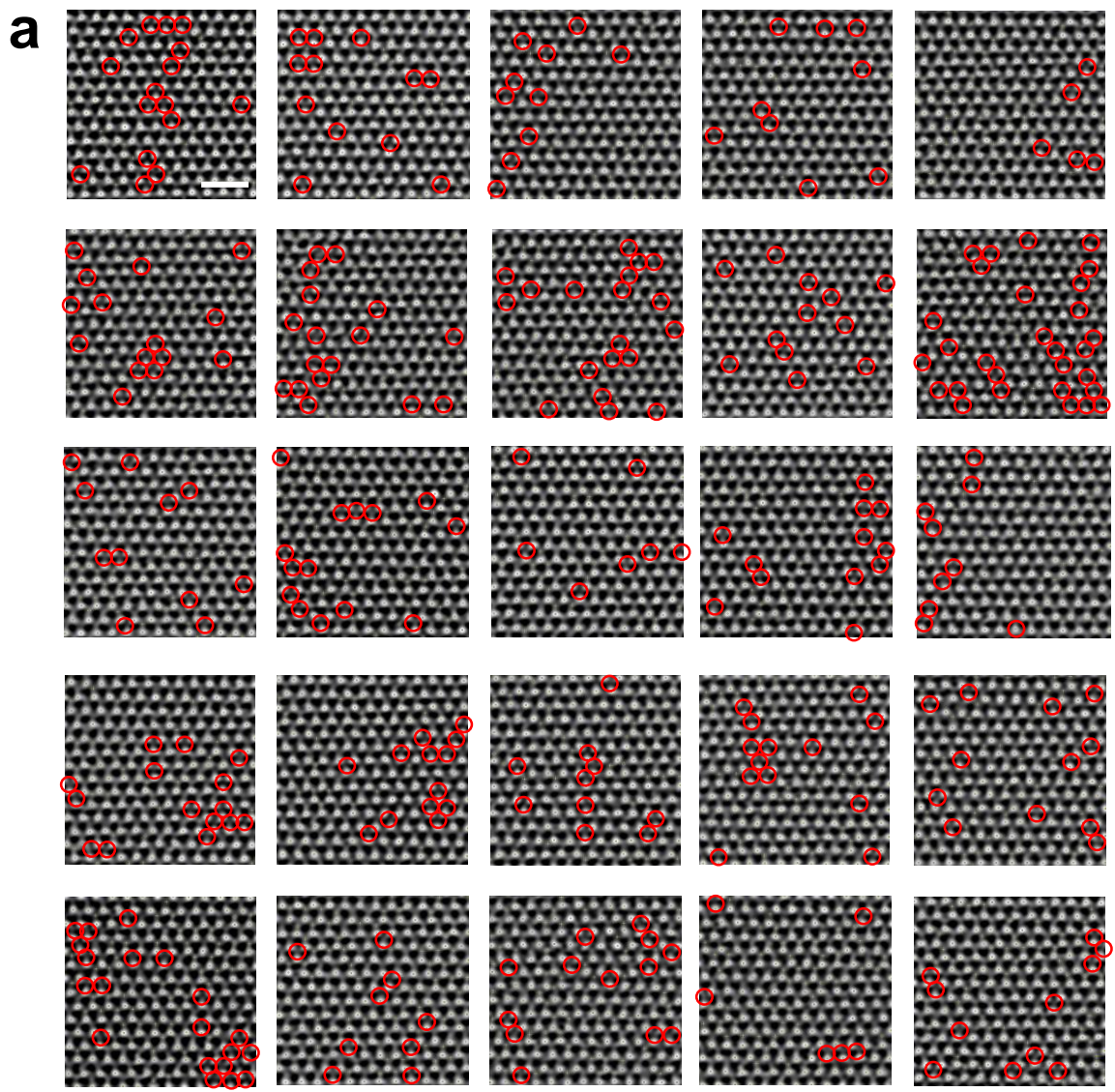

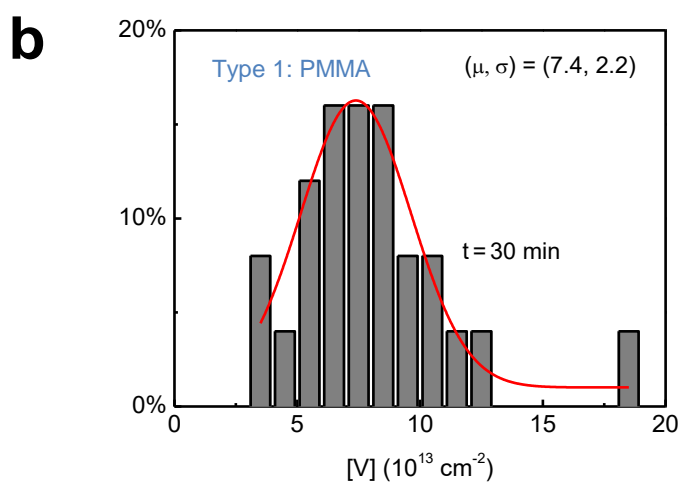

**Figure S4. Statistics of [V] for a 30-min defect engineered MoS$_2$ on hydrophobic PMMA surfaces (low vacancy generation rate). a**, Atomically resolved STEM images for 25 independent areas. The sulfur vacancies are denoted by red circles. Scale bar: 1 nm. **b**, Histogram of [V] showing an average value μ ~7.4×10$^{13}$ cm$^{-2}$ and a standard deviation σ ~2.2×10$^{13}$ cm$^{-2}$.

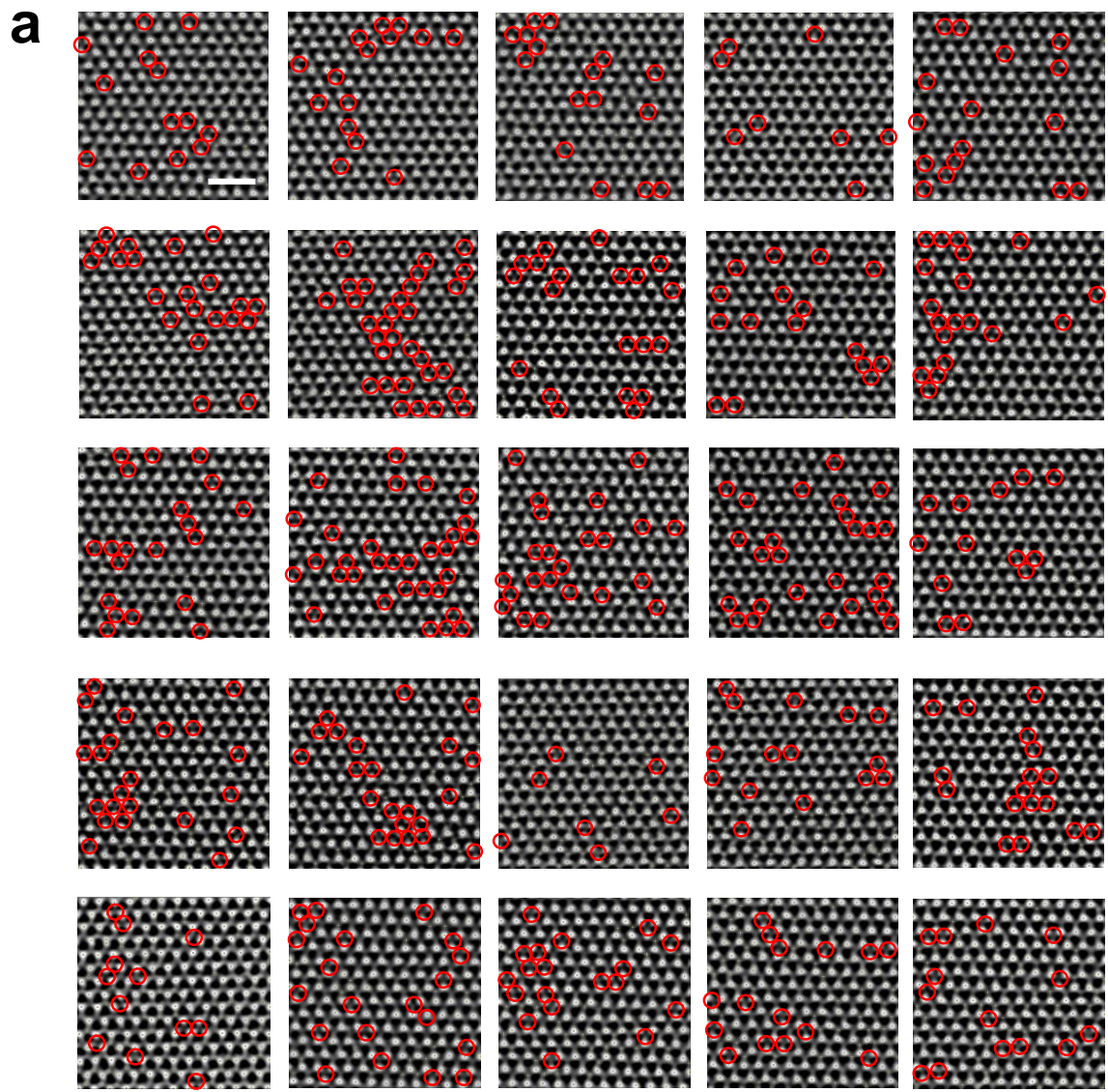

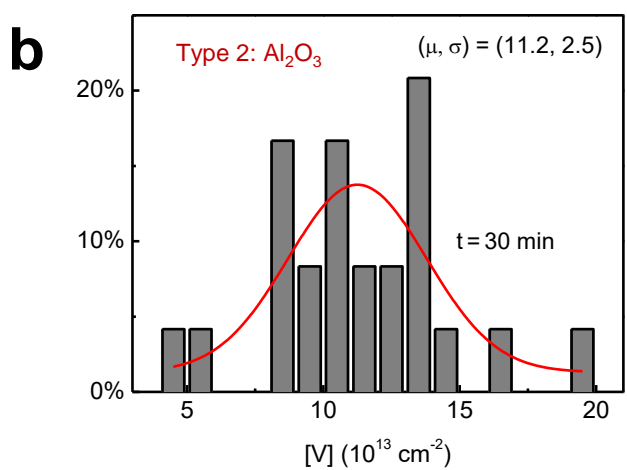

**Figure S5. Statistics of [V] for a 30-min defect engineered MoS$_2$ on hydrophilic Al$_2$O$_3$ surfaces (high vacancy generation rate). a**, Atomically resolved STEM images for 25 independent areas. The sulfur vacancies are denoted by red circles. Scale bar: 1 nm. **b**, Histogram of [V] showing an average value μ ~11.2×10$^{13}$ cm$^{-2}$ and a standard deviation σ ~2.5×10$^{13}$ cm$^{-2}$.